\documentclass[aps,prd,showpacs,showkeys,nofootinbib,superscriptaddress,preprintnumbersonecolumn,12pt]{revtex4-1}

\usepackage[utf8]{inputenc}
\usepackage{graphicx}
\usepackage{amsfonts}
\usepackage{amssymb}
\usepackage{amsmath}
\usepackage{setspace}
\usepackage{booktabs}
\usepackage[shellescape,latex]{gmp}
\usempxpackage{amssymb}

\def\bb{\mathbb}
\def\sqr#1#2{{\vcenter{\hrule height.#2pt
   \hbox{\vrule width.#2pt height#1pt \kern#1pt
      \vrule width.#2pt}
   \hrule height.#2pt}}}

\def\bsqr#1#2{{\vrule width #1pt height#2pt}}
\def\bsquare{{\mathchoice\bsqr66\bsqr66\bsqr33\bsqr33}}
\def\badbreak{\penalty1000}

\def\R{{\bb R}}				    
\def\fir{{\scriptscriptstyle{\text{\rm IR}}}}                     
\def\fuv{{\scriptscriptstyle{\text{\rm UV}}}}                     
\def\cro{{\scriptscriptstyle{\text{\rm A}}}}                      
\def\lm0{{\lambda_0}}                                             
\def\lmm{{\lambda_m}}                                             

\newcommand*{\GtrSim}{\smallrel\gtrsim}
\newcommand*{\LessSim}{\smallrel\lesssim}
\newcommand*{\GtrApprox}{\smallrel\gtrapprox}
\newcommand*{\LessApprox}{\smallrel\lessapprox}

\makeatletter
\newcommand*{\smallrel}[2][.8]{%
  \mathrel{\mathpalette{\smallrel@{#1}}{#2}}%
}
\newcommand*{\smallrel@}[3]{%
  \sbox0{$#2\vcenter{}$}%
  \dimen@=\ht0 %
  \raise\dimen@\hbox{%
    \scalebox{#1}{%
      \raise-\dimen@\hbox{$#2#3\m@th$}%
    }%
  }%
}
\makeatother

\def\fm {\mathop{\hbox{fm}}}

\usepackage{amsmath} 
\usepackage{dsfont}  
\usepackage{bm}      

\def\beq{\begin{equation}}
\def\eeq{\end{equation}}
\def\beqs#1\eeqs{\beq\begin{split} #1 \end{split}\eeq}

\long\def\comment#1{}

\usepackage{microtype}
\usepackage[dvipsnames]{xcolor}
\usepackage[colorlinks=true,backref=false, linktocpage=true,
citecolor=PineGreen,urlcolor=PineGreen,linkcolor=PineGreen,pdfpagemode=UseOutlines]{hyperref}

\hypersetup{%
  bookmarksnumbered=true,
  pdftitle = {},
  pdfsubject = {},
  pdfauthor = {},
  pdfkeywords = {}
}

\begin{document}

\title{Possible New Phase of Thermal QCD}

\author{Andrei\ Alexandru}
\email{aalexan@gwu.edu}
\affiliation{The George Washington University, Washington, DC 20052, USA}

\author{Ivan Horv\'ath}
\email{ihorv2@g.uky.edu}
\affiliation{University of Kentucky, Lexington, KY 40506, USA}

\date{Nov 22, 2019}

\begin{abstract}

Using lattice simulations, we show that there is a phase of thermal QCD, where 
the spectral density $\rho(\lambda)$ of Dirac operator changes as $1/\lambda$ for 
the infrared eigenvalues $\lambda<T$. This behavior persists over the entire low energy 
band we can resolve accurately, over 3 orders of magnitude on our largest volumes. 
We propose that in this ``IR phase", the well-known non-interacting scale invariance 
at very short distances (UV, $\lambda \rightarrow \infty$, asymptotic freedom), 
coexists with very different interacting type of scale invariance at long distances 
(IR, $\lambda<T$). Such dynamics may be responsible for the unusual fluidity 
properties of the medium observed at RHIC and LHC. We point out its connection 
to the physics of Banks-Zaks fixed point, leading to the possibility of massless 
glueballs in the fluid. Our results lead to the classification of thermal QCD phases 
in terms of IR scale invariance. The ensuing picture naturally subsumes the standard 
chiral crossover feature at $``T_c" \approx 155$ MeV. Its crucial new aspect is 
the existence of temperature $T_\fir$ (200 MeV $< T_\fir < $ 250~MeV) marking 
the onset of IR phase and possibly a true phase transition. 

\end{abstract}



\maketitle


\noindent
{\bf 1.~Introduction. $\;$}The study of strongly interacting matter as a function of temperature 
and baryon density is an active area of theoretical and experimental research 
(see~\cite{Busza:2018rrf} for recent review). At high 
energies of colliding heavy nuclei, such as those studied at LHC and the high end of RHIC, baryon 
densities are small enough so that the results are generally expected, among other things, to shed 
light on the nature and properties of thermal QCD transition in the early universe. In this regime, 
it has become widely accepted, largely due to the matured power of lattice QCD~\cite{Aoki:2006we}, 
that increasing temperature leads to a smooth crossover in properties of thermal strongly 
interacting matter. On the experimental side, results from 
RHIC~\cite{Arsene:2004fa,Back:2004je,Adams:2005dq,Adcox:2004mh}  and LHC~\cite{Muller:2012zq} 
based on modeling the time evolution of collisions in terms of relativistic hydrodynamics, 
produced a picture of a strongly coupled liquid-like medium with extremely low $\eta/s$ 
(shear viscosity/entropy density) at high temperatures. In parallel and initially independent 
developments, similar values of $\eta/s$ were obtained in highly symmetric and strongly coupled 
gauge theories with large number of colors, studied by means of their holographic 
dual~\cite{Policastro:2001yc}. This sparked a flurry of attempts to model the medium seen 
in the experiments via more refined descriptions of this type. 

However, the physics of thermal QCD transition(s) and the nature of the discovered liquid-like 
state of matter are far from settled, even in the limit of vanishing net baryon density 
($\mu \!=\!0$), the setting of our interest. Among other things, the currently favored scenario 
involving a single feature (crossover at``$T_c$") offers limited room for accommodating the 
dramatic change from a medium described as a weakly interacting hadron resonance gas 
to a strongly interacting near-perfect fluid. 
In this work, we propose a hierarchy of thermal effects in QCD, based on scale invariance 
properties at long distances, which adds new detail to the existing picture and ties with 
it in a natural manner. Special role in our analysis will be played by glue fields. In fact, one 
of our conclusions is that, from the standpoint of scale invariance, the phase structure of pure 
glue SU(3) gauge theory (pgQCD) and that of nature's strong interactions (QCD) are qualitatively 
the same. Avoiding the complication of quark fields, we thus first describe the proposed picture 
in the context of the former.

Since pgQCD is a theory of massless vector fields, it is classically scale invariant. According 
to the standard picture at zero temperature, this scale invariance is broken by quantum effects, 
leading to low energy scale and the spectrum of massive bound states (glueballs). 
Yet, at asymptotically short distances, the system can be effectively described by 
perturbing non-interacting gluons (asymptotic freedom~\cite{Gross:1973id,Politzer:1973fx}). 
This is sometimes rephrased as scale invariance being broken at long distances (IR), but present 
at asymptotically short distances (UV) in the trivial non-interacting form.

Here we propose and support the following behavior of thermal pgQCD. Turning the temperature 
gradually on, the scale (non)invariance properties of a thermal state remain similar 
to that of a zero-temperature vacuum, until the scale of thermal agitation becomes 
comparable to the lowest scale of broken scale invariance (``gluon condensate"). This is 
characterized by the crossover temperature $T_\cro$ past which the properties of thermal 
medium change rapidly toward the restoration of scale invariance in IR. The latter then 
occurs at a well-defined temperature $T_\fir > T_\cro$. In the ensuing range 
$T_\fir < T < T_\fuv$ (IR phase), gauge fields characteristic of a thermal state are scale invariant 
at distances larger than $\approx 1/T$.
Unlike asymptotic scale invariance in UV, present at all temperatures, IR invariance emerges 
due to the interaction that is still strong at long distances. 
For $T > T_\fuv$ (UV phase), the field fluctuations in IR regime ($\lambda < T$)  
effectively disappear, and the notion of IR scale invariance becomes trivial. 
The system can then be described as a weakly interacting gluon plasma.

\begin{figure}[t]
\begin{center}
    \centerline{
    \hskip 0.1in
    \includegraphics[width=14.0truecm,angle=0]{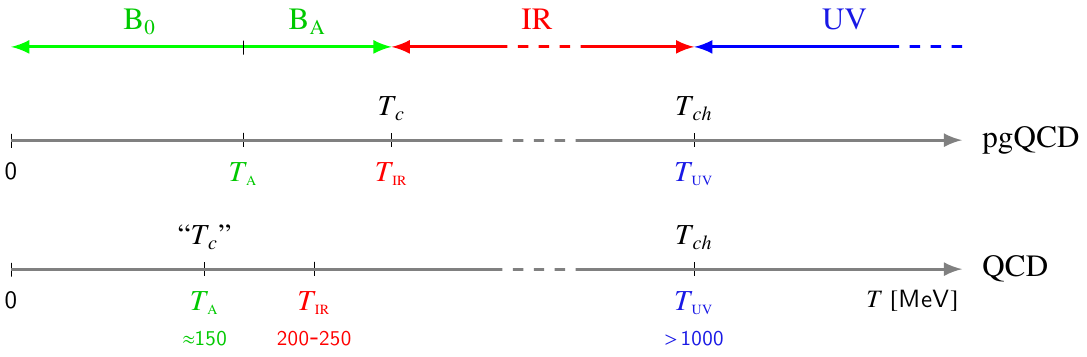}
    }
    \vskip -0.05in
    \caption{Common thermal phase structure of pure glue QCD (pgQCD) and QCD in terms of 
     scale invariance. Since pgQCD is but a model of QCD glue, setting its physical scales 
     involves a small arbitrariness. Temperatures in black appeared in literature without 
     reference to scale invariance.}
    \label{fig:phase1}
    \vskip -0.40in
\end{center}
\end{figure}

This scenario is schematically shown in Fig.~\ref{fig:phase1} (top, middle). Note that 
the low-temperature region $T<T_\fir$ (B phase for ``broken") is split into two regimes 
B$_0$ and B$_\cro$ by $T_\cro$. The relation to transition temperatures discussed 
previously in literature without invoking scale invariance is also indicated. 
Temperature $T_\fir$ coincides with the well-known $T_c$ of Polyakov line first order 
transition in pgQCD~\cite{Svetitsky:1982gs}. 
In addition, we identify $T_\fuv$ with $T_{ch}$ of chiral polarization
transition~\cite{Alexandru:2012sd,Alexandru:2014paa,Alexandru:2015fxa}.
Analog of $T_\cro$ has not appeared in the context of 
pgQCD.\footnote{Transition with analogous physical meaning was in fact discussed 
in Ref.~\cite{Alexandru:2015fxa} but, rather than being attributed to a distinct dynamical 
effect, it was mistakenly identified with $T_c$ in pgQCD.}

\vfill\eject

Next, we present evidence that the above $T$-pattern of scale invariance in gauge 
field is retained by QCD of nature (Fig.~\ref{fig:phase1}, bottom). In other words, 
scaling properties of QCD glue, which again enters as nominally scale-free entity, 
are driven by quantum nature of the theory like in pgQCD, rather than quark mass 
effects. Thus, there is a crossover temperature $T_\cro \!\approx\! 150\,$MeV which we 
qualitatively associate with chiral ``$T_c$" of the standard scenario. However, here it 
is simply a characteristic temperature of B phase, marking the onset of changes toward 
IR scale invariance. Note that the quark condensate now also plays a role in determining 
the value of $T_\cro$. The IR phase then emerges at $200 < T_\fir < 250\,$MeV. 

\noindent
Before proceeding to lattice evidence, we address several immediate questions.

\smallskip

\noindent 
$(i)$ Since lattice offers good quantitative control over QCD at $\mu=0$, how did the IR 
phase escape the detection? The answer is insufficient volumes. Indeed, the usual expectation 
is that IR scales $\Lambda < T$ contribute little to physics for $T >``T_c"$. Our proposal 
not only contradicts this but implies that, for $T_\fir < T < T_\fuv$, it is the deep 
infrared scales $\Lambda \ll T_\fir$ that drive a significant IR contribution. Hence, 
we predict the existence of a ``crossover size" $L_\fir(T) \gg 1/T_\fir$ of the system 
past which the deep infrared physics becomes readily reflected in thermal observables. 
The systems of sufficiently large spatial sizes $L > L_\fir(T)$ are not commonly studied 
at present. This is expanded upon in Appendix~\ref{app:cic}.

\smallskip

\noindent
$(ii)$ Given $(i)$, how is the existence of IR phase inferred from lattice simulations? 
At $T \GtrSim T_\fir$, we detect the onset of scale invariant $1/\lambda$ behavior of 
Dirac spectral density $\rho(\lambda)$ (number of eigenmodes per unit volume and spectral 
interval) for $\lambda \LessSim T$. We propose that this arises due to the onset 
of effective IR scale invariance of glue fields dominating the thermal 
state.\footnote{The strict claim 
$\rho(\lambda) \propto 1/\lambda$ for $\lambda \LessSim T$ seemingly entails integrability
issues, but these are superficial in light of regularizations involved. For this discussion, 
one can simply assume $\rho(\lambda) \propto 1/\lambda^{1-\delta}$, $0 < \delta \ll 1$.} 
While the two notions are not equivalent apriori, they are consistent 
(Appendix~\ref{app:si}). Moreover, in theories with IR scale invariant gauge fields, such 
as those governed by Banks-Zaks fixed point, the pure power law behavior of $\rho(\lambda)$ 
is expected due to its proposed connection to mass anomalous 
dimension~\cite{Patella:2012da,Cheng:2013eu}.\footnote{Note that 
$\rho(\lambda) \propto 1/\lambda^{1-\delta}$,  $0 < \delta \ll 1$, would imply very 
large anomalous dimension in those theories.} 
This argument also suggests that, up to small quark mass deformations, IR scale invariance 
of glue extends to quark sector in QCD, which was implicitly assumed already.

\smallskip

\noindent
$(iii)$ Scale invariance in field theory is normally addressed via the energy-momentum 
tensor. However, such test should only include the scales up to 
$\Lambda_\fir(T) \LessSim T$ (upper edge of $1/\lambda$) in this case. 
In conjunction with $(i)$, this avoids the conflict with existing lattice 
results~\cite{Borsanyi:2013bia,Bazavov:2014pvz}.

\smallskip

\noindent
$(iv)$ Given its perturbative nature, the UV phase should only ensue when thermal 
agitation mostly engages perturbative scales. In that vein, our expectation is that 
$T_\fuv > 1\,$GeV (Fig.~\ref{fig:phase1}). Its precise determination in lattice 
simulations is challenging in part because the minimal system size needed to detect 
the IR phase grows with temperature (Appendix~\ref{app:cic}). 

\begin{figure}[t]
\begin{center}
    \centerline{
    \hskip 0.00in
    \includegraphics[width=5.0truecm,angle=0]{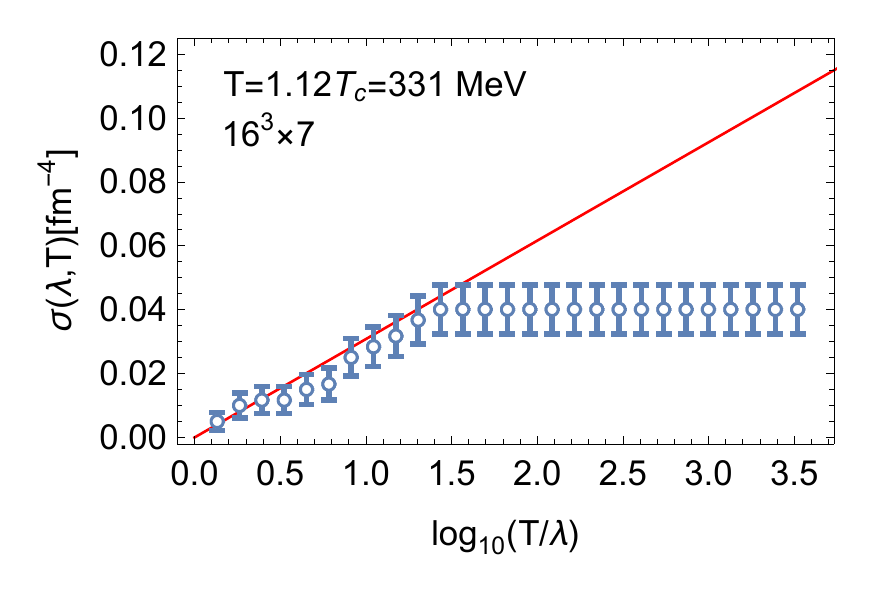}
    \hskip -0.11in
    \includegraphics[width=5.0truecm,angle=0]{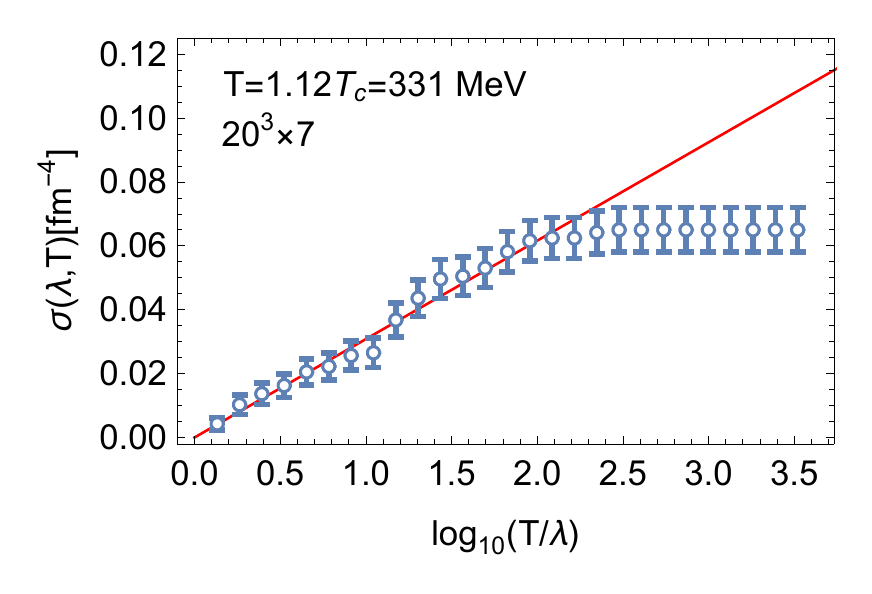}
    \hskip -0.11in
    \includegraphics[width=5.0truecm,angle=0]{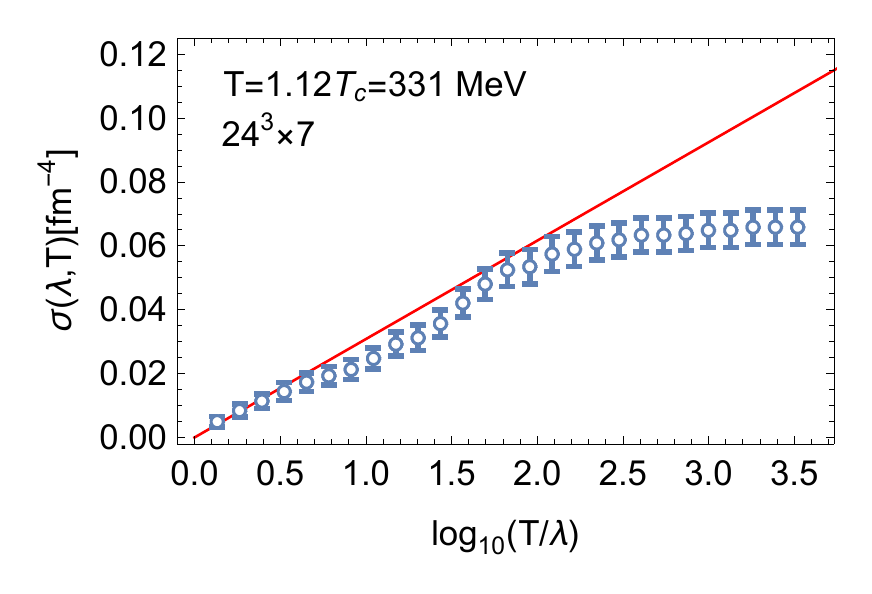}
    }
     \vskip -0.14in
    \centerline{
    \hskip 0.00in
    \includegraphics[width=5.0truecm,angle=0]{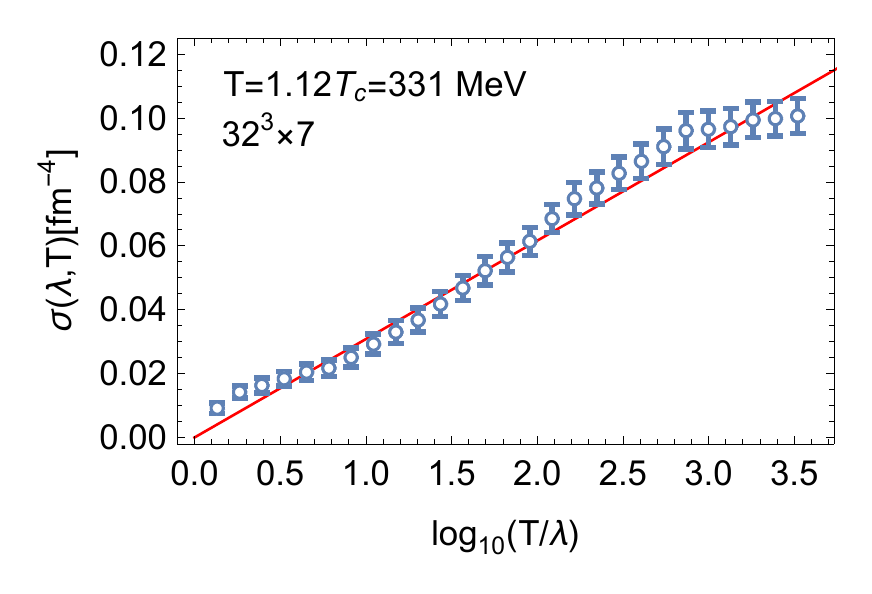}
    \hskip -0.11in
    \includegraphics[width=5.0truecm,angle=0]{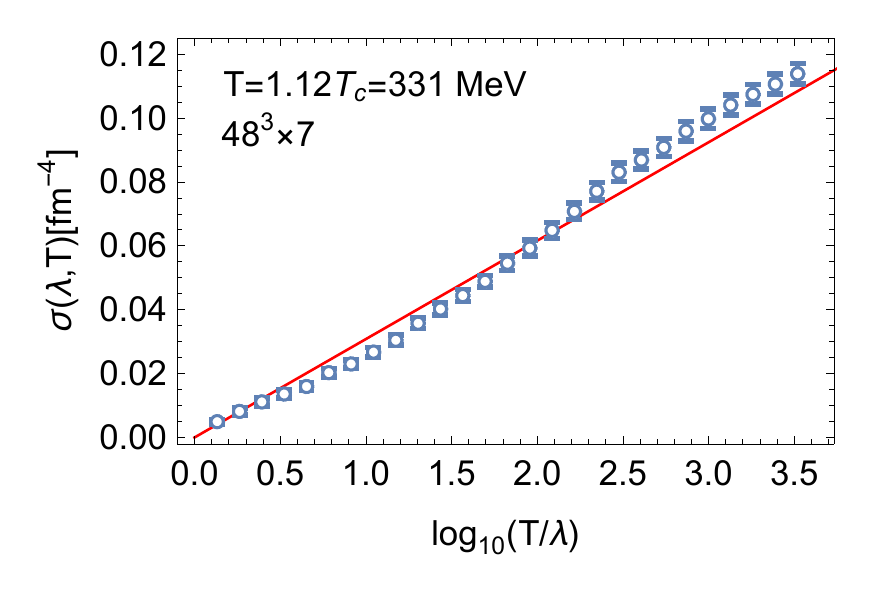}
    \hskip -0.11in
    \includegraphics[width=5.0truecm,angle=0]{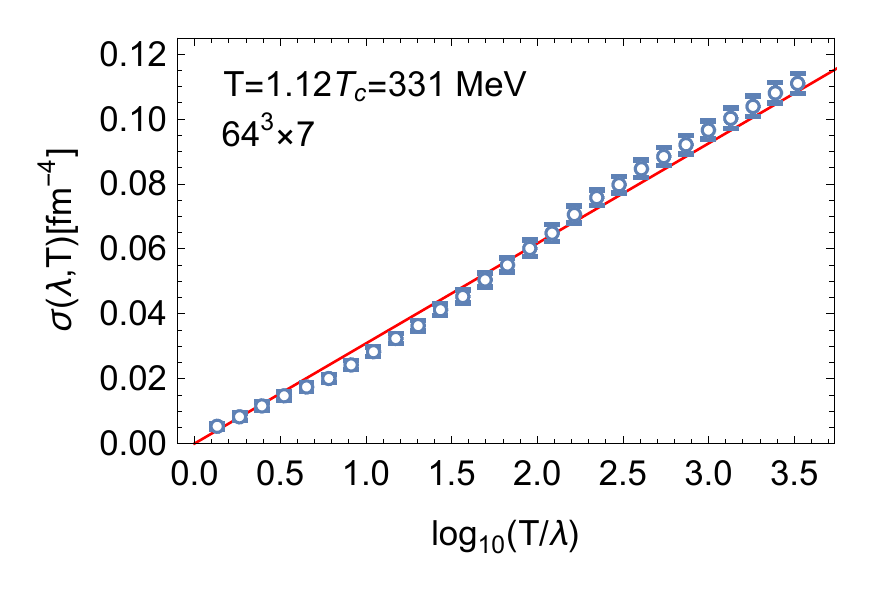}
     }
     \vskip -0.18in
     \caption{IR dependence $\rho(\lambda) \propto 1/\lambda$ emerging in pgQCD 
     at $T=1.12\, T_c$ and UV cutoff $a=0.085\fm$.}
     \label{fig:infrared}
    \vskip -0.40in
\end{center}
\end{figure} 

\medskip

\noindent
{\bf 2. Lattice Evidence. $\;$} Technical details of our simulations are summarized 
in Appendix~\ref{app:td}. To discuss the results, we start with pgQCD where
needed volumes are more readily accessible. In Ref.~\cite{Edwards:1999zm}, a peak 
at the infrared end of Euclidean Dirac spectral density has been observed in 
pgQCD above $T_c$. Only recently it was shown~\cite{Alexandru:2015fxa} that this 
feature is not a regularization artifact. Here we present evidence that 
$\rho(\lambda) \propto \lambda^{-1}$ in IR which, together with 
$\rho(\lambda) \propto \lambda^3$ in UV, generates a bimodal structure facilitating 
scale invariance at both ends of the spectrum.

To that end, we study the spectrum of the overlap Dirac operator on equilibrium 
backgrounds. A useful quantifier is the volume density of eigenmodes in spectral 
range $[\lambda,T]$, namely
\begin{equation}
    \sigma(\lambda,T) \equiv \int_\lambda^T \rho(\omega) \,d\omega
    \qquad \longrightarrow \qquad c(T) \,\ln \frac{T}{\lambda} 
    \quad\; \text{for} \quad\; \rho(\lambda) = \frac{c}{\lambda}    \quad
    \label{eq:020}
\end{equation}
If $\rho(\lambda) \propto \lambda^{-1}$ for $\lambda < T$, a straight line passing 
through the origin is obtained in variable $x = \ln T/\lambda \ge 0$. Note that 
$\lambda \!=\! T$ corresponds to $x\!=\!0$ and IR is approached by increasing $x$. 
If $\rho(\lambda) \propto \lambda^{-1}$ only for $\lambda < \Lambda_\fir(T) < T$, 
a y-shifted linear segment appears for $x > \ln T/\Lambda_\fir$. 

In Fig.~\ref{fig:infrared} we show $\sigma(x)$ in pgQCD on increasing volumes 
(UV cutoff $a\!=\!0.085$ fm) at $T\!=\!1.12\,T_c$. Each case involves an easily 
identifiable, approximately linear segment extending from origin to increasingly 
IR scales as the IR cutoff $L$ increases. Leveling off at larger $x$ 
signals the IR edge of the spectrum. On the largest volume 
($L=5.4\,$fm), the $1/\lambda$ behavior persists over 3.5 orders of magnitude 
from $T$ down to deep infrared. The eye-guiding red line is the same 
in each plot and corresponds to the fit for largest volume, producing
the value $c(1.12\,T_c)=0.0308(3)\,$fm$^{-4}$ in Eq.~\eqref{eq:020}. 
In Appendix~\ref{app:ad}, we discuss a more direct approach to exposing
the $1/\lambda$ dependence of $\rho(\lambda)$ over wide range of scales.

\begin{figure}[t]
\begin{center}
    \centerline{
    \hskip 0.00in
    \includegraphics[width=5.0truecm,angle=0]{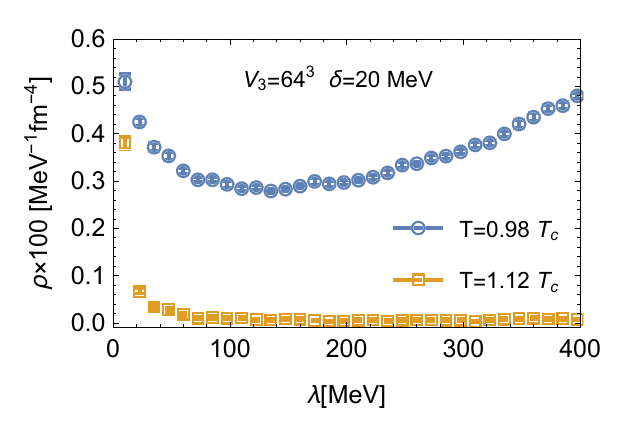}
    \hskip -0.08in
    \includegraphics[width=5.0truecm,angle=0]{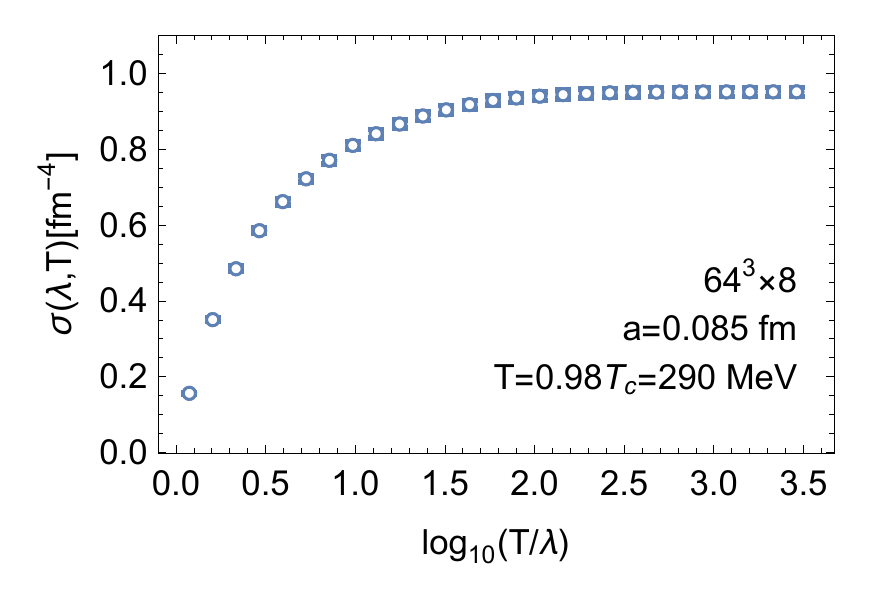}
    \hskip -0.08in
    \includegraphics[width=5.0truecm,angle=0]{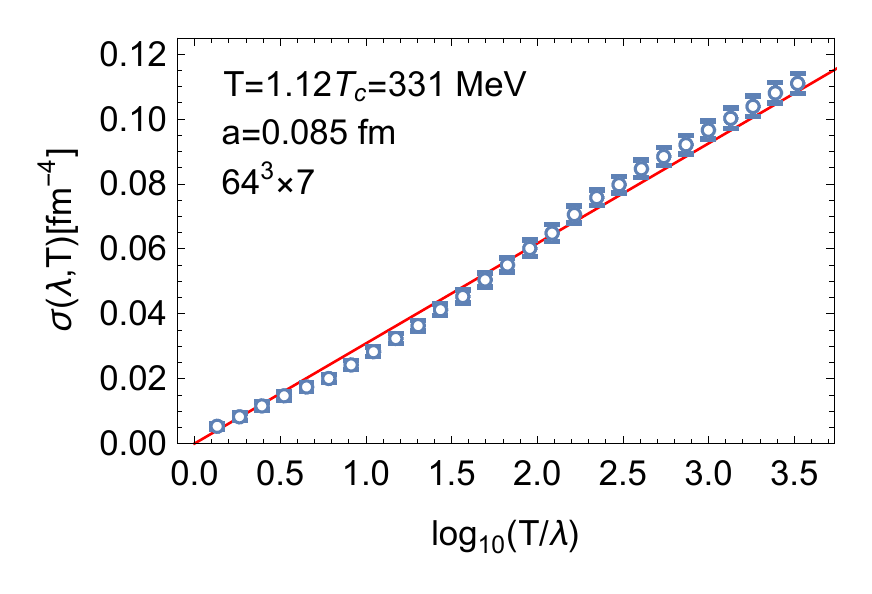}
    }
    \vskip -0.12in
    \centerline{
    \hskip 0.00in
    \includegraphics[width=5.0truecm,angle=0]{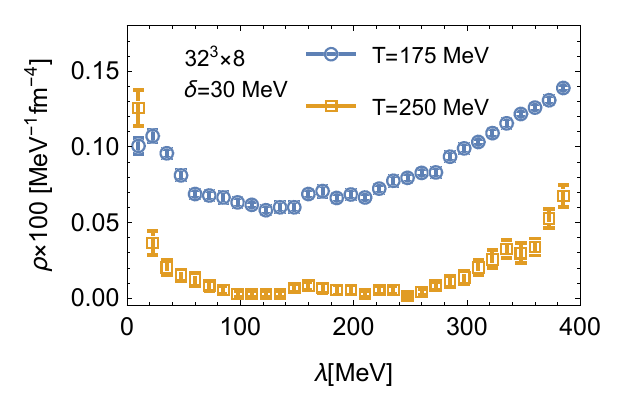}
    \hskip -0.08in
    \includegraphics[width=5.0truecm,angle=0]{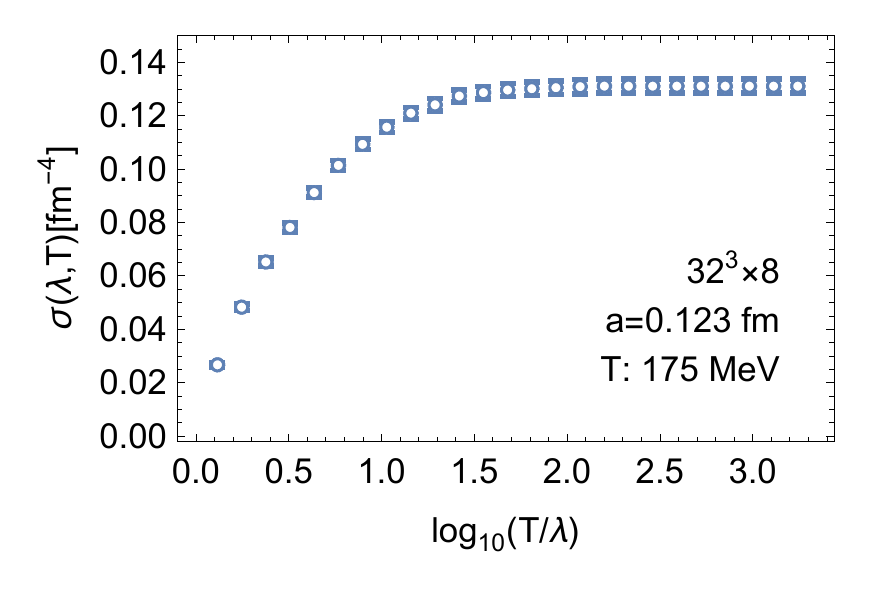}
    \hskip -0.08in
    \includegraphics[width=5.0truecm,angle=0]{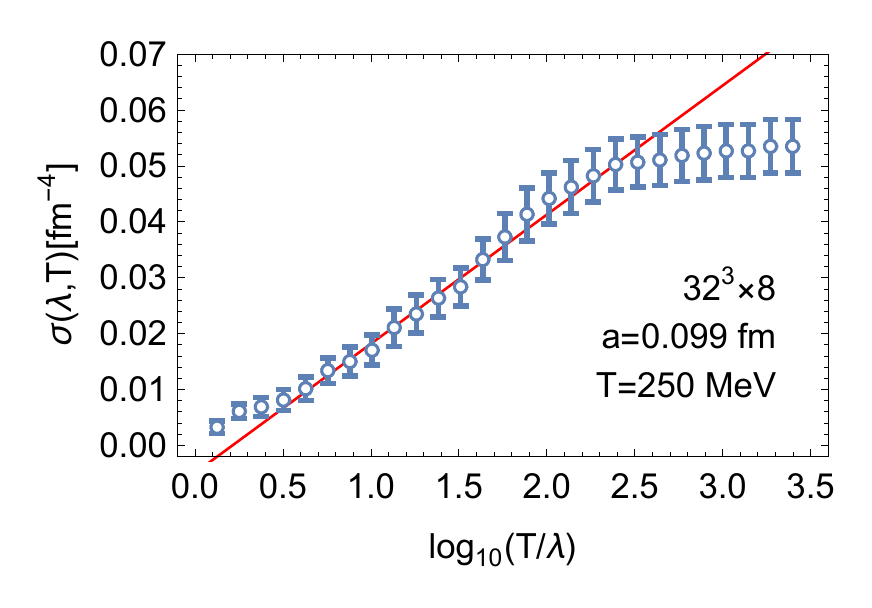}
    }
    \vskip -0.25in
    \caption{Thermal transition to IR phase in pure glue QCD (top) and QCD (bottom).}
    \label{fig:IRcross}
    \vskip -0.50in
\end{center}
\end{figure}

To assess the relationship of Polyakov line phase transition in pgQCD to its IR 
phase, we simulate the system at $T \!=\! 0.98\, T_c$ in the otherwise identical 
setup with large volume. The resulting $\rho(\lambda)$ is shown in 
Fig.~\ref{fig:IRcross} (top left). Apart from saturation at the IR edge of 
the spectrum, we find no linear segment in the corresponding $\sigma(x)$ 
(top middle), in direct contrast to $1/\lambda$ behavior at $T\!=\!1.12\, T_c$ 
(top right). Thus, barely below $T_c$, the system is in the B phase. Note 
also the characteristic difference in spectral densities between B 
and IR phases (top left). Given the above and the corroborating spectral 
evidence of Ref.~\cite{Alexandru:2015fxa} at $T \!=\! 1.02 \,T_c$, we conclude 
that $T_\fir$ coincides with $T_c$. One consequence of this is that B and IR 
phase of pgQCD are separated by a first order phase transition.

Important feature of the Dirac spectrum at $T\!=\!0.98 \,T_c$ is that $\rho(\lambda)$ 
exhibits the IR peak even at $T < T\fir$. Indeed, there is a minimum of 
$\rho(\lambda)$ at $\lmm \!\approx\! 120\,$MeV (Fig.~\ref{fig:IRcross}, top left). 
Such minimum at $\lmm > 0$ may exist even at zero temperature due to the
possible logarithmic divergence at $\lambda \to 0$ and/or the presence of positive 
power with negative prefactor~\cite{Osborn:1998qb}. However, this has not yet been 
confirmed in pgQCD simulations, implying that $\lmm(T)$ is very small 
or zero at low $T$. 
This leads us to propose that it is meaningful to distinguish the $T \GtrSim 0$ 
and $T \LessSim T_\fir$ regimes by a crossover characterized by temperature 
$0 < T_\cro < T_\fir$. While the crossover point is a non-unique concept, here we have 
in mind a commonly used approach based on the rate of change. In other words, 
we define $T_\cro$ as the position of maximum (peak) in $d\lmm/dT$.
In physics terms, $T_\cro$ relates to the point at which gluon condensate becomes 
significantly affected by thermal agitation. It splits the B phase 
into regimes B$_0$ and B$_\cro$ (Fig.~\ref{fig:phase1}) with the latter referred 
to as anomalous, conforming to terminology of Ref.~\cite{Alexandru:2015fxa}. 

Standard expectations suggest that the IR phase, commencing at $T_\fir$, ends at temperature 
$T_\fuv$ (Fig.~\ref{fig:phase1}) above which scales $\lambda \!\approx\! T$ become amenable 
to perturbative treatment. Since IR peak is not featured in weakly coupled regime, we 
define $T_\fuv$ as a temperature at which $\rho(\lambda)$ becomes a nondecreasing function 
on $\lambda \ge 0$ with $\rho(0)=0$. The associated disappearance of IR peak has been observed 
on moderate volumes in Refs.~\cite{Alexandru:2012sd,Alexandru:2014paa}, accompanied 
by the simultaneous loss of chiral polarization in low-lying Dirac modes. Since the latter effect is 
characterized by temperature $T_{ch}$, we propose that $T_\fuv = T_{ch}$ as indicated in 
Fig.~\ref{fig:phase1}.

We now turn to overlap spectral densities in QCD. More specifically, we study SU(3) 
gauge theory with $N_f=2+1$ quark flavors at physical masses 
(see Appendix~\ref{app:td}), which is a very precise representation of real-world 
strong interactions. To support the existence of $T_\fir$, we show in 
Fig.~\ref{fig:IRcross} (bottom) the analog of IR transition we described in pgQCD. 
The clearly noninvariant behavior at $T\!=\!175\,$MeV is contrasted with that at 
$T\!=\!250\,$MeV. The latter exhibits characteristic features of the IR phase, both 
in terms of $\rho(\lambda)$ and $\sigma(x)$. Regarding the latter, note also 
the similarity to the pattern displayed by volume sequence in Fig.~\ref{fig:infrared}. 
In Appendix~\ref{app:ad} we present additional results at $T\!=\!200\,$MeV, featuring
the behavior more marginally on the B$_\cro$ side. This leads us to the following 
initial estimates 
\begin{equation}
    200\,\text{MeV} \,<\, T_\fir \,<\, 250\,\text{MeV}  \qquad ,\qquad
    T_\cro \,\approx\, 150\,\text{MeV} \LessApprox ``T_c"
    \label{eq:030}
\end{equation}
where $``T_c"$ is the temperature of chiral crossover. 
The estimate of $T_\cro$ follows from our analysis in Ref.~\cite{Alexandru:2015fxa} 
(Fig.~5 of that work). There it was found that $\lmm$ becomes essentially undetectable 
at $T\!=\!150\,$MeV (simulated size $L = 5.3\,$fm) signaling either its large 
drop at lower temperatures, so that significantly larger volumes are needed to reveal 
it, or its entire disappearance ($\lmm \to 0$). Both options entail the proximity of 
a dividing point between B$_0$ and B$_\cro$ regimes, namely $T_\cro$.  


\vfill\eject

\noindent
{\bf 3. IR-UV Separation and Banks-Zaks Fixed Point.}
The signature aspect of transition at  $T_\fir$ is a clean separation of IR and UV scales 
in the gauge field, reflected by almost perfectly bimodal $\rho(\lambda)$. 
Additional data illustrating the latter is presented in Appendix~\ref{app:ad}. The analysis of  
Refs.~\cite{Alexandru:2014paa,Alexandru:2014zna,Alexandru:2015fxa,Alexandru:2015shf}
revealed that, apart from increasing the temperature, such IR-UV separation is also inducible 
by decreasing the quark mass or increasing the number of flavors in parameter space 
($T,m_i,N_f$) of SU(3) gauge theories with fundamental quarks. Our aim is to integrate 
the new element of IR scale invariance into these findings, which promises a valuable 
insight into the nature of IR phase in thermal QCD. 

We start in the corner of SU(3) theory space which is native to IR scale invariance, 
namely the vicinity of conformal window~\cite{Banks:1981nn} 
($N_f$ massless flavors, $N_f^c < N_f < 16.5$, $T\!=\!0$). 
In Refs.~\cite{Alexandru:2015fxa,Alexandru:2014paa,Alexandru:2014zna,Alexandru:2015shf}
it was found that small mass dynamics at $N_f \!=\! 12$, believed by most researchers to 
be  near-conformal, generates the pattern of IR-UV separation closely mimicking that of QCD 
in the IR phase. While originally interpreted as indicating an unexpectedly large $N^c_f$, 
the revelation that $\rho(\lambda)$ may be a pure power in IR begs this to be reconsidered
since $\rho(\lambda) \propto \lambda^p$ is exactly what one expects near conformality. 
A consistent inference is the split of conformal window into a strongly coupled part 
$N_f^c \equiv N_f^\fir < N_f  < N_f^\fuv$ with $p<0$, and a weakly coupled part 
$N_f^\fuv < N_f \le 16.5$ with $p>0$. The parametric trends in IR-UV separation 
then lead us to propose that strongly coupled regimes $T_\fir < T < T_\fuv$ 
of QCD and $N_f^\fir < N_f  < N_f^\fuv$ of the conformal window belong to a single 
contiguous IR phase in SU(3) theory space, defined by $p<0$. In this sense, 
the observed elements of IR scale invariance in thermal QCD descend 
from conformality of a strongly coupled Banks-Zaks fixed point.

The above argument introduces an unconventional scenario for dynamics in a strongly
coupled conformal window which requires more detail. Consider the $T\!=\!0$ system 
at $N_f^\fir < N_f < N_f^\fuv$ with varying degenerate quark mass $m$. At large $m$, 
glue behaves as in pgQCD: there is no scale invariance in IR and the gauge coupling 
runs indefinitely. At asymptotically small $m$, the running stops at dynamically 
generated $\Lambda_\fir>0$, and scale invariance ensues at larger distances: 
the interacting fixed point entirely runs the IR, while the Gaussian one governs UV. 
The evidence of
Refs.~\cite{Alexandru:2014paa,Alexandru:2014zna,Alexandru:2015fxa,Alexandru:2015shf}
suggests that Banks-Zaks fixed point turns on in a sudden manner by theory entering 
the IR phase ($m < m_\fir$) in mass. In more detail, the sequence
\begin{equation}
   m = \infty \quad \xrightarrow{\quad \text{B}_0 \quad} \quad
   m_\cro \quad \xrightarrow{\quad \text{B}_\cro \quad}  \quad 
   m_\fir \quad \xrightarrow{\quad \text{IR} \quad} \quad 0
   \label{eq:040}
\end{equation}
produces B and IR phases as in the thermal case. With $N_f$ in a strongly coupled 
regime, the UV phase does not materialize. Thus, the mass vicinity of a theory in 
strongly coupled  conformal window (its IR phase) is characterized by 
$\rho(\lambda) \propto \lambda^p$ where\footnote{Note that conformality 
constraints on unitarity~\cite{Mack:1975je} and the conjectured 
method~\cite{Patella:2012da} of extracting $\bar{\psi} \psi$ anomalous 
dimension then raise interesting questions on details of $m \to 0$ limit and 
its relation to $m \equiv 0$.}
\begin{equation}
    p < 0 
    \quad\;\; , \quad\;\;  0 < m < m_\fir 
    \quad\;\; , \quad\;\;  0 < \lambda < \Lambda_\fir 
    \quad\;\; , \quad\;\;  \lim_{m \to 0} \Lambda_\fir(m) > 0 \quad\;\,
    \label{eq:050}
\end{equation}

Among appealing features of the above is that it naturally explains the generation 
of $\Lambda_\fir$. Indeed, the degree of IR-UV separation reflects the extent 
to which IR and UV fixed points in Wilson theory space dominate the dynamics of 
given field theory in respective scale domains. The nearly perfect scale separation 
in the IR phase suggests that these domains are described by almost decoupled IR 
and UV subtheories. 
Since $\Lambda_\fir$ defines the domain of IR, integrating out UV degrees of freedom 
(UV theory) only feeds into the dynamics of IR (IR theory) in a trivial manner. 
This stops the running since IR theory is scale invariant.

To see the relevance of this connection, consider the $m \to 0$ 
limit in SU(3) theory space.\footnote{We refer to theories defined as
$\lim_{m \to 0}\, \lim_{L \to \infty}$ of those with $N_f$ mass-degenerate flavors.}
For $N_f\!=\!2$ (chiral QCD) at $T\!=\!0$, this produces massive physical 
spectrum except for Goldstone pions. In contrast, for $N_f^\fir < N_f < N_f^\fuv$ it 
leads to massless behavior in all channels due to the approached conformality. 
Since the property of infinite correlation length is stable within the contiguous 
IR phase,\footnote{Due to the above monotonicity properties,  the IR phase in this 
restriction remains contiguous.} it extends from the strongly coupled conformal regime 
($N_f^\fir < N_f < N_f^\fuv$,  $T=0$) to the IR regime of chiral QCD  
($N_f=2$, $T_\fir < T < T_\fuv$). Our approach thus predicts that the lowest 
``screening masses" (spatial correlations) and ``quasiparticle masses" 
(time-like correlations) are zero in the IR phase of chiral QCD. Note that we do not 
claim the absence of excitations with masses larger than $T$. In fact, additive 
contributions to correlators by masses larger than $\Lambda_\fir(T,N_f)$ distinguish 
the low 
energy scale invariance of IR phase from strict conformality present only at $T=0$. 

Strong interactions of nature turn on small $m$ at $N_f=2$, but the above picture of 
IR phase only gets corrected by light lowest masses in meson and baryon correlation 
functions. The IR scale invariance of glue, inherent to IR phase, is expected to keep 
correlators of glue operators long-range, and the associated glueball-like excitations 
massless.

\vfill\eject

\noindent
{\bf 4. Synthesis and Main Points. $\;$}
We proposed the existence of a new phase in thermal QCD, the IR phase 
$T_\fir < T < T_\fuv$, 
featuring aspects of scale invariance at distances larger 
than $1/\Lambda_\fir$, where $\Lambda_\fir(T) \LessSim T$. In particular, 
our way of probing the system suggests that glue fields dominating 
the thermal state 
in the IR phase 
are statistically self-similar 
upon rescalings involving such distances (Appendix~\ref{app:si}). 

In the standard scenario, QCD matter enters the near-perfect fluid regime 
above the chiral crossover temperature $``T_c" \!\approx\! 155\, \text{MeV}$. 
However, given that scale invariance underlies model descriptions able 
to mimic the observed fluidity properties~\cite{Policastro:2001yc}, we 
propose that this transition actually occurs at $T_\fir$ 
($200 < T_\fir < 250\, \text{MeV}$). In other words, the strongly 
interacting near-perfect fluid is realized by the IR phase. If glue fields 
continue to follow the described patterns arbitrarily deep into IR, then 
$T_\fir$ marks a phase transition where the leading IR power 
in $\rho(\lambda)$ changes from $p=0$ to $p \GtrApprox -1$. 
This transition could be consequential for the analysis of heavy ion 
experiments and for modeling the thermal history of the universe.

The observed elements of IR scale invariance can be understood 
by viewing thermal QCD in the larger context of asymptotically 
free SU(3) gauge theories with fundamental quarks. To that end, we 
proposed the phase structure in this space that can be summarized by 
\vskip -0.15in
\begin{singlespace}
\begin{equation}
     \text{phase}  \;=\;
     \begin{cases} 
        \;\;\text{B}    &     \text{if} \quad  p=0  \\[2pt]
        \;\;\text{IR}   &     \text{if} \quad  p<0  \\[2pt]
        \;\;\text{UV}  &     \text{if} \quad  p>0        
     \end{cases}
     \qquad  \text{where}  \qquad 
     \rho(\lambda)   \,\propto\,   \lambda^p   \quad\; \text{for} \quad\;    
     \lambda \to 0 \quad
\end{equation}
\end{singlespace}
\noindent with transitions occurring accordingly. For example, increasing 
the temperature past $T_\fir$ in QCD is expected to eventually generate 
a transition from $p \GtrApprox -1$ to $p > 1$, identifying
$T_\fuv$.\footnote{
The value of $p$ in thermal UV phase could be infinite if the depletion 
of modes in the infrared proceeds faster than arbitrary positive power,
e.g. if gap develops in the Dirac spectrum. Note also that $p=0$
(B phase) includes the case of logarithmically diverging density.}
The connection to scale invariance stems from the proposed existence of
a contiguous IR phase in SU(3) space, in which UV field fluctuations 
associated with Gaussian fixed point 
($A^\fuv ,\, \rho \propto \lambda^3$) coexist with IR fluctuations driven 
by strongly coupled Banks-Zaks fixed point 
($A^\fir ,\,\rho \propto \lambda^{-1+\delta}$). 
The high $T$ ($T_\fir < T < T_\fuv$) medium of QCD (near-perfect fluid) 
and low $T$ ($T \GtrSim 0$) medium of a strongly coupled near-conformal
theory ($N_f^\fir < N_f < N_f^\fuv$) both belong to this phase and thus 
share important dynamical features. This may prove useful in 
guiding the analytic attempts to describe the near-perfect fluid.

The conjecture that IR phase of QCD realizes the near-perfect fluid 
is expected to have phenomenological consequences. For example, 
using the above connection to the physics of strongly coupled conformal 
window, we have argued for highly unusual spectrum of excitations 
(quasiparticles and screening masses) in the fluid.
The fluctuations originating from a strongly coupled Banks-Zaks fixed 
point generate a narrow IR band of massless (glueball-like) and 
light (hadron-like) excitations. At the same time, fluctuations tied to 
Gaussian fixed point produce excitations in the UV band, creating 
a large gap ($>\!T$). We hypothesize that the remarkable IR-UV 
separation, both in terms of fluctuating quantum fields and the resulting 
physical excitations, is one of the key ingredients driving the exotic 
properties of the strongly-coupled medium discovered at the RHIC and 
LHC.

The central point of this work, namely the existence of a new infrared 
dynamics in gauge theories (IR phase), invites additional corroboration 
and further clarification. For example, a large scale study confirming  
$\rho(\lambda) \propto 1/\lambda$ in QCD at the level comparable 
to pgQCD (Fig.~\ref{fig:infrared}) is of primary interest. Similar 
quantitative analyses in the vicinity of a strongly coupled conformal 
window (e.g. $N_f \!=\!12$) are also very desirable. Studies examining 
the proposed association of power law Dirac densities with scale 
invariant gauge fields by direct means would solidify and refine 
the interpretation of the IR phase. In the process, such works may also 
clarify why $1/\lambda$ (rather than other power) emerges in QCD. 

While this manuscript was in preparation, work~\cite{Rohrhofer:2019qwq} 
appeared, suggesting a different new feature in thermal QCD based
on chiral considerations.

\begin{acknowledgments}
A.A. is supported in part by the National Science Foundation CAREER 
grant PHY-1151648 and by U.S. DOE Grant No. DE-FG02-95ER40907.
I.H. acknowledges the support from Department of Anesthesiology at 
the University of Kentucky, help from Jian Liang and Robert Mendris, 
as well as conversations with Anatoly Dymarsky and Ganpathy Murthy.
We thank the Wuppertal-Budapest collaboration for sharing their lattice 
ensembles. 
\end{acknowledgments}

\begin{appendix}

\section{The Crossover Size}
\label{app:cic}

The IR phase of QCD is characterized by $\rho(\lambda) \propto 1/\lambda$ for 
$\lambda < \Lambda_\fir$. Here $\Lambda_\fir$ determines the minimal distance
over which scale invariance of glue fields applies. But what is the minimal size 
of the system in which this behavior affects an observable in a discernible 
manner? This role is played by the ``crossover size" 
$L_\fir$ (Sec.~1).\footnote{$L_\fir$ can be viewed as a size at which 
the associated finite volume correction assumes its asymptotic form.} 
Given that the IR contribution is driven by deep infrared ($\ll \Lambda_\fir$) 
rather than the vicinity of $\Lambda_\fir$, it is clear that 
$L_\fir \gg 1/\Lambda_\fir$ for standard observables. Moreover, the density 
of Dirac eigenmodes in the IR regime drops quickly with temperature 
(see e.g. \cite{Alexandru:2014paa}), causing $L_\fir$ to increase. Sensitivity 
to scale invariant behavior of glue is then expected on systems 
of size $L$ satisfying 
\begin{equation}
    L(T) \,>\, L_\fir(T) \,>\, L_\fir(T_\fir) \,\gg\, 1/\Lambda_\fir(T_\fir) 
    \,\GtrSim\, 1/T_\fir
    \label{eq:cic05}
\end{equation}
where the last inequality is due to $\Lambda_\fir(T) \LessSim T$. Hence, 
$L(T) \gg 1/T_\fir$ applies to all standard observables and all temperatures 
$T_\fir < T < T_\fuv$. Since $T_\fir$ is comparable to $\Lambda_{QCD}$, 
the relevant sizes are larger than typically considered sufficient for thermal 
QCD studies. 

Lattice introduces a slight complication in that the Dirac operator, serving 
as the detector of IR scale invariance, is not a unique object: different 
discretizations capture aspects of continuum behavior to varying degree. 
Chirality plays a relevant role here. Indeed, the bimodality in 
$\rho(\lambda)$ was first observed with overlap operator that fully respects 
chirality, while it was not seen by the staggered operator on identical 
backgrounds~\cite{Edwards:1999zm}. However, the IR peak has recently been 
identified by staggered-type operator in pgQCD on larger 
volumes~\cite{Kovacs:2017uiz}, confirming that the presence of this feature 
is discretization independent. This is also consistent with bimodality 
of the overlap operator persisting into the continuum limit, shown 
in Ref.~\cite{Alexandru:2015fxa}. In addition, since $L_\fir$ has physical 
origin (see above), we expect that
\begin{equation}
     \lim_{a \to 0} L_\fir^{lat_1}(T,a) \,=\, 
     \lim_{a \to 0} L_\fir^{lat_2}(T,a) \,=\, L_\fir(T) 
     \label{eq:cic10}    
\end{equation}
i.e. that $L_\fir(T)$ is universal for fixed definition of the crossover point.  

In lattice QCD, Dirac operator defining the quark part of the action obviously plays 
a special role. While the existence of IR peak in this ``native" Dirac spectrum 
appears more difficult to ascertain numerically, the studies focusing on 
the U$_A$(1) problem~\cite{Tomiya:2016jwr,Sharma:2018syt} already suggest that 
the feature is present at physical light-quark masses, albeit the studied volumes are 
small. Its absence would in fact be very surprising. Indeed, the ensuing singularity 
in the space of lattice Dirac operators with respect to~\eqref{eq:cic10}, as well as 
the associated possibility of non-universality in topological susceptibility, make 
such scenario unlikely. 

\section{Scale Invariance and Dirac Spectral Density}
\label{app:si}

Our aim here is to illustrate how scale invariance of gauge field $A$ constrains the form of 
spectral density $\rho(\lambda \mid A)$ of Euclidean Dirac operator $D=D[A]$. This is easiest to do 
in $\R^4$, the setup relevant for theories in conformal window, but the arguments can be modified
to finite temperature. We implicitly assume that $A_\mu(x) \in su(N)$ although this is not important 
in the present context. Thus, we are dealing with eigenvalue problem on fixed ``classical" 
background, defined by ($A_\mu(x)$ is anti-Hermitian)
\begin{equation}
    D[A] \psi(x) \,\equiv\, \sum_\mu \gamma_\mu
    \Bigl[\, \frac{\partial}{\partial x_\mu} - A_\mu(x) \, \Bigr] \psi (x)
    \,=\, i \, \lambda \, \psi(x)
    \label{eq:si10}
\end{equation}
where $\lambda \in \R$ and $\psi$ is an eigenmode. Let $A^{(s)}$ be the gauge field obtained 
from $A$ by the canonical scale transformation. The following are the simultaneous eigensystem 
triples
\begin{equation}
    (A, \psi, \lambda)      \quad \longleftrightarrow \quad (A^{(s)}, \psi^{(s)}, s \lambda)
    \qquad\quad A^{(s)}(x) \,\equiv\, s A(sx) 
    \quad , \quad \psi^{(s)} \propto \psi(sx)
    \quad 
    \label{eq:si20}
\end{equation}
where the correspondence is one-to-one. Envisioning the potentials singular at origin or infinity, we 
consider the regularized eigenvalue problem on $[\epsilon,L]^4$ with $\epsilon$ the ultraviolet and $L$
the infrared regulator.\footnote{The more symmetric setup on 
$(\, [-L/2,-\epsilon/2] \cup [\epsilon/2,L/2] \,)^4$ proceeds in an analogous way.} 
The relation \eqref{eq:si20} is then modified as
\begin{equation}
    (A, \psi, \lambda, \epsilon, L)      \quad \longleftrightarrow \quad 
    (A^{(s)}, \psi^{(s)}, s \lambda, \epsilon/s, L/s)
    \label{eq:si30}
\end{equation}
The standard (anti)periodic boundary conditions on $A$, $\psi$ are respected by the correspondence.

With the usual assumption that the spectrum on finite volume is discrete, \eqref{eq:si30} implies that 
the number of eigenmodes in interval $[\lambda_1, \lambda_2]$ for setup on the left is the same as 
that in $[s\lambda_1, s\lambda_2]$ for setup on the right. Focusing on $A(x)$ with no singularity at 
$x \to \infty$ allows us to remove infrared cutoff ($L \to \infty$) and to account for number of 
eigenmodes in terms of smooth spectral density. This then leads to
\begin{equation}
    \int_{\lambda_1}^{\lambda_2} d \lambda\, \rho(\lambda \mid A, \epsilon) =
    \frac{1}{s^4} \int_{s\lambda_1}^{s\lambda_2} d \lambda\, \rho(\lambda \mid A^{(s)}, \epsilon/s) =
    \frac{1}{s^3} \int_{\lambda_1}^{\lambda_2} d \lambda\, \rho(s\lambda \mid A^{(s)}, \epsilon/s)
    \label{eq:si40}
\end{equation}
for all $\lambda_1$ and $\lambda_2$. Consequently,
\begin{equation}
    s^3 \rho(\lambda \mid A, \epsilon) = 
    \rho(s\lambda \mid A^{(s)}, \epsilon/s)
    \label{eq:si50}
\end{equation}
which for scale invariant background $A^{(s)}(x) = A(x)$ leads to
\begin{equation}
    \rho(\lambda, \epsilon) = \lambda^3 f(\lambda \epsilon)
    \label{eq:si60}
\end{equation}
where $f(x)$ is an arbitrary non-negative function. Thus, for scale invariant free field 
($A(x) \equiv 0$), with no singularity at the origin, the density is $\epsilon$-independent 
and $\rho(\lambda) \propto \lambda^3$. However, no leading infrared power, such as the behavior 
$1/(\lambda \epsilon^4)$, is excluded a priori. In quantum theory, the diverging UV cutoff length 
scale is replaced by the dynamically generated $1/\Lambda_\fir$, and we thus have 
$\rho(\lambda) \propto \Lambda_\fir^4/\lambda$. These considerations can be generalized to 
self-similar (rather than strictly scale invariant) gauge backgrounds, providing additional freedom 
to accommodate the $1/\lambda$ dependence. 

\section{Summary of Technical Details}
\label{app:td}

Our pgQCD simulations were performed using Wilson gauge action with scale setting based 
on the reference value $r_0\!=\!0.5\,$fm. The volume dependence of Dirac spectra was studied 
at $\beta\!=\!6.054$ which corresponds to UV cutoff $a\!=\!0.085\,$fm. The estimate of 
$T_c$ involved the results of Ref.~\cite{Necco:2003vh}. The ensembles at $T=1.12\,T_c$
($N_\tau\!=\!7$) contain 400,400,400,200,200,100 gauge configurations for 
$N\!=\!16,20,24,32,48,64$ systems respectively ($1/T=N_\tau a$ and $L=N a$). The results 
at $T=0.98\,T_c$ ($N_\tau=8,\, N=64\,$) are based on 94 configurations.

For QCD Dirac spectra, we utilized the gauge ensembles of Wuppertal-Budapest group described
in Ref.~\cite{Borsanyi:2010bp}. More precisely, they were generated in $N_f=2+1$ theory at 
physical light quark mass of $(m_u+m_d)/2$, and the physical ``heavy" quark mass of $m_s$. 
In terms of lattice setup, the simulations used tree-level Symanzik-improved gauge action and 
stout-improved staggered fermions. The physical point (thus scale setting) was defined by 
fixing $m_\pi$, $m_K$ and $f_K$ to their physical values at zero temperature. Our analysis
is based on 100 gauge configurations in each case.

\begin{figure}[t]
\begin{center}
    \centerline{
    \hskip 0.00in
    \includegraphics[width=6.0truecm,angle=0]{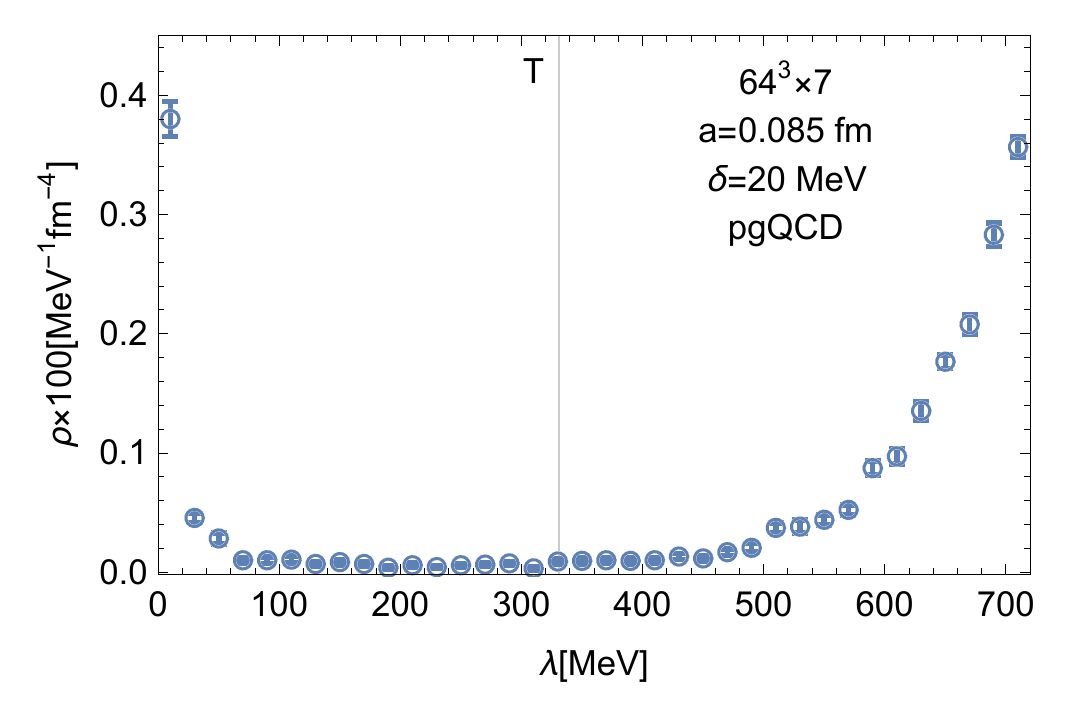}
    \hskip 0.1in
    \includegraphics[width=6.0truecm,angle=0]{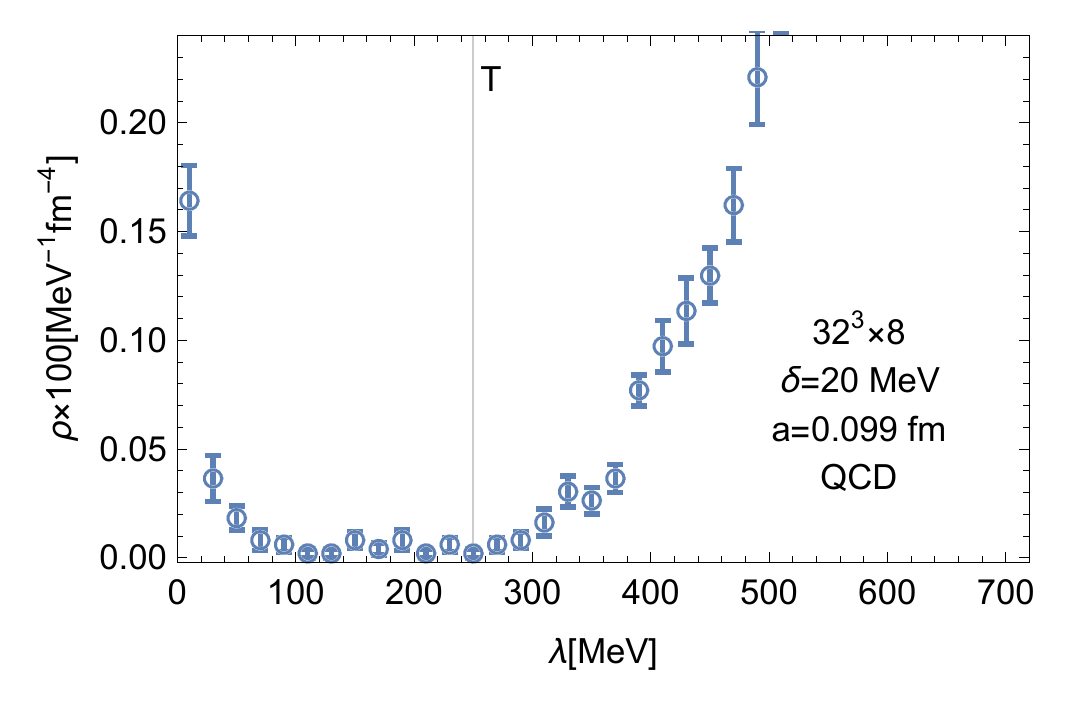}
    }
    \vskip -0.1in
    \caption{The separation of IR and UV scales, manifested in a sharply bimodal overlap 
    Dirac spectral density in pgQCD at $T\!=\!1.12 T_c$ (left) and QCD at $T\!=\!250\,$MeV.}
    \label{fig:double}
    \vskip -0.40in
\end{center}
\end{figure}

The main object of our interest is the 4-volume density $\sigma(\lambda_1,\lambda_2)$ of 
Dirac eigenmodes from spectral interval $[\lambda_1, \lambda_2]$ (convention set by 
Eq.~\eqref{eq:si10}). This quantity is commonly expressed in terms of the corresponding 
spectral density $\rho(\lambda)$, namely
\begin{equation}
     \sigma(\lambda_1,\lambda_2) \,\equiv\, 
     \int_{\lambda_1}^{\lambda_2} d \lambda \, \rho(\lambda) 
     \label{eq:td10}         
\end{equation}
Unless stated otherwise, exact zero modes are excluded from counting. On finite 4-volume 
$L^3/T$, the ensemble average is implicitly assumed in \eqref{eq:td10}, although expressing 
$\rho(\lambda)$ in terms of $\delta$-functions makes it meaningful even for a single 
configuration. 

In a numerical study, it is necessary to work with coarse-grained version of 
$\rho(\lambda)$. This is achieved by introducing the parameter $\delta>0$ and defining
\begin{equation}
    \rho(\lambda,\delta) \,\equiv\, 
    \frac{\sigma(\lambda-\delta/2, \lambda+\delta/2)}{\delta}
    \qquad , \qquad
    \rho(\lambda) \,=\, \lim_{\delta \to 0} \rho(\lambda,\delta)
    \label{eq:150}              
\end{equation}
Only $|\lambda|>\delta/2 + \epsilon$ with suitably chosen $\epsilon > 0$ to avoid finite 
volume effects is shown or quoted in any given $\rho(\lambda,\delta)$. 
A Wilson-Dirac based overlap 
operator~\cite{Neuberger:1997fp} with parameters $\rho=26/19$ and $r=1$ was used in all 
Dirac spectrum calculations.

Implicitly restarted Arnoldi method~\cite{Sorensen:1992fk,Lehoucq:1996} was used to compute 
the eigenvalues and eigenvectors of the overlap operator. For all but one ensemble 
used in this study, it is efficient to first compute the eigenvalues of $D^\dagger D$ in 
a chiral sector, and then reconstruct the eigenvalues of $D$ using standard 
techniques~\cite{Alexandru:2011sc}. For the $N=64$ pgQCD lattice at $T=1.12 T_c$, 
it becomes problematic to distinguish the eigenvalues of near-zero eigenmodes from those 
of exact zero modes. To ensure the reliability of numerical results in this case, we 
solved the eigenvalue problem for $D$ directly, utilizing a suitable polynomial spectral 
transformation to accelerate the convergence.

\begin{figure}[t]
\begin{center}
    \centerline{
    \hskip 0.00in
    \includegraphics[width=5.7truecm,angle=0]{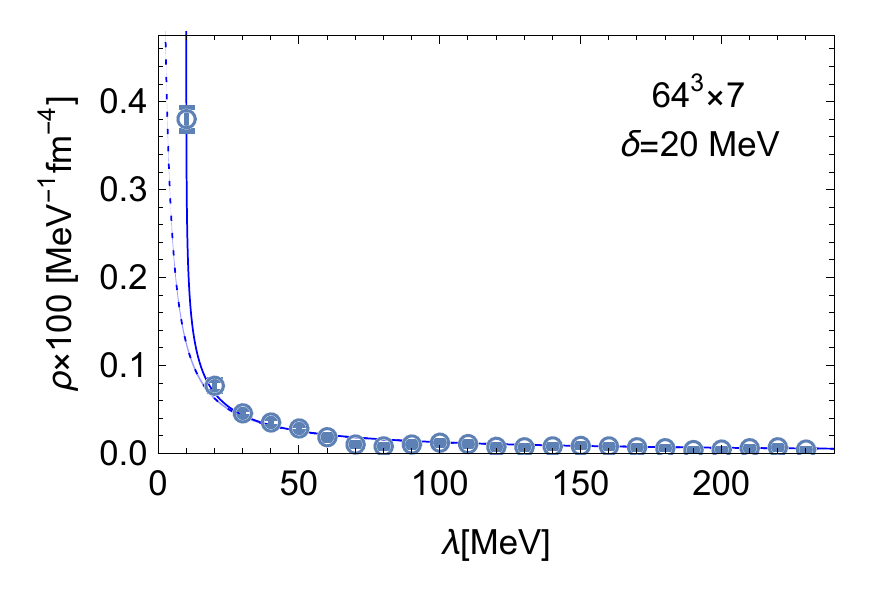}
    \hskip -0.11in
    \includegraphics[width=5.7truecm,angle=0]{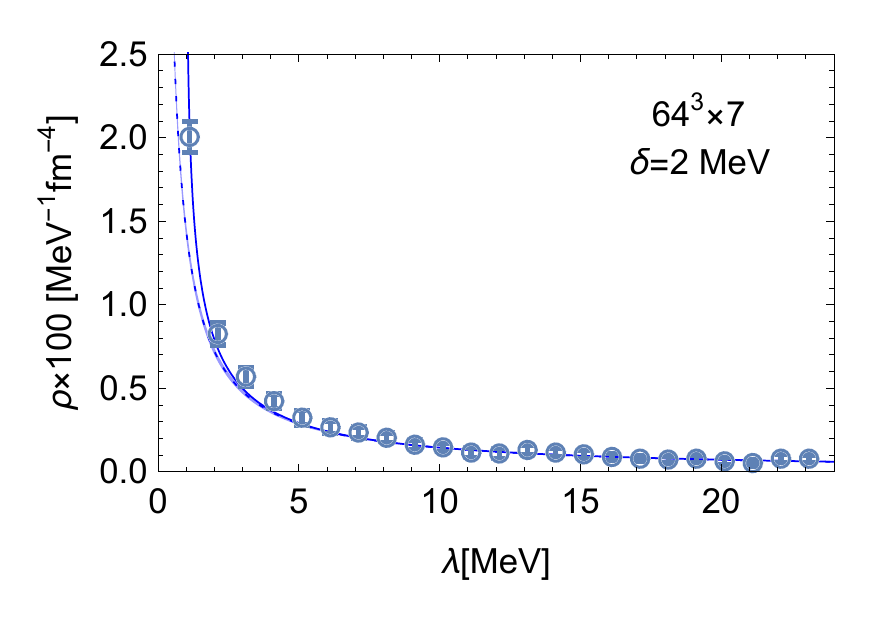}
    \hskip -0.11in
    \includegraphics[width=5.5truecm,angle=0]{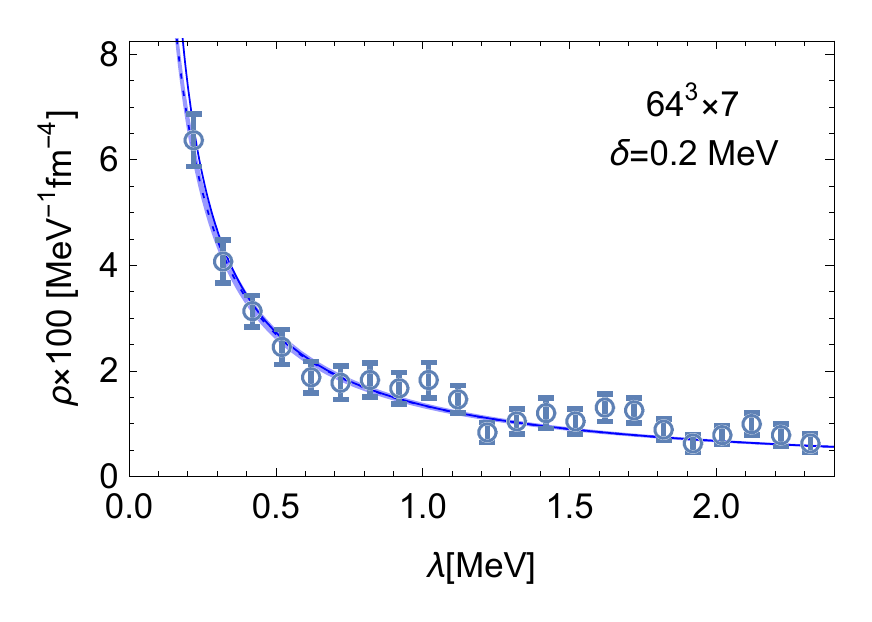}
     }
     \vskip 0.0in
     \caption{Overlap Dirac spectral density for pgQCD in IR phase ($T\!=\!1.12\, T_c$) over 
     spectral ranges (and bin sizes) scaled by factors of 10. Dashed lines represent 
     direct $1/\lambda$ fits, while the solid lines include the correction on finite bin 
     size.}
     \label{fig:sizoom}
    \vskip -0.40in
\end{center}
\end{figure}

\section{Additional Data}
\label{app:ad}

In this Appendix we present additional lattice data to further support our 
conclusions.

A key to the proposed picture of thermal phases is the emergence of a remarkable 
separation of IR and UV physics at $T_\fir$. This signature aspect of IR phase 
is reflected in the sharp bimodality of Dirac spectral density and the resulting clear 
separation of scales (Sec.~3). To convey this feature explicitly, we show in 
Fig.~\ref{fig:double} spectral densities for both pgQCD and QCD in the IR phase. 
The data suggests the presence of dynamics in which IR and UV regimes act as 
separate independent ``components" of the theory. 

The $1/\lambda$ behavior of $\rho(\lambda)$ over wide IR range of scales can 
also be checked in a more direct manner, namely by the process of zooming in 
toward the infrared. In Fig.~\ref{fig:sizoom} we show this for pgQCD at 
$T=1.12\, T_c$ on our largest lattice. With the lower spectral edge fixed at 
$0.1\,$MeV to avoid finite volume effects, we plot $\rho(\lambda)$ up to 
240, 24 and 2.4 MeV respectively, with bin sizes correspondingly scaled. In 
each case we fit the data to $1/\lambda$ with bin size taken into account in 
the procedure (solid line) to avoid the finite bin distortion.  

\begin{figure}[t]
\begin{center}
    \centerline{
    \hskip 0.00in
    \includegraphics[width=5.0truecm,angle=0]{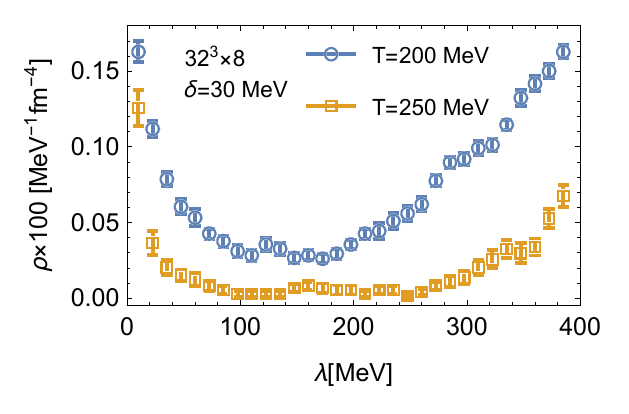}
    \hskip -0.08in
    \includegraphics[width=5.0truecm,angle=0]{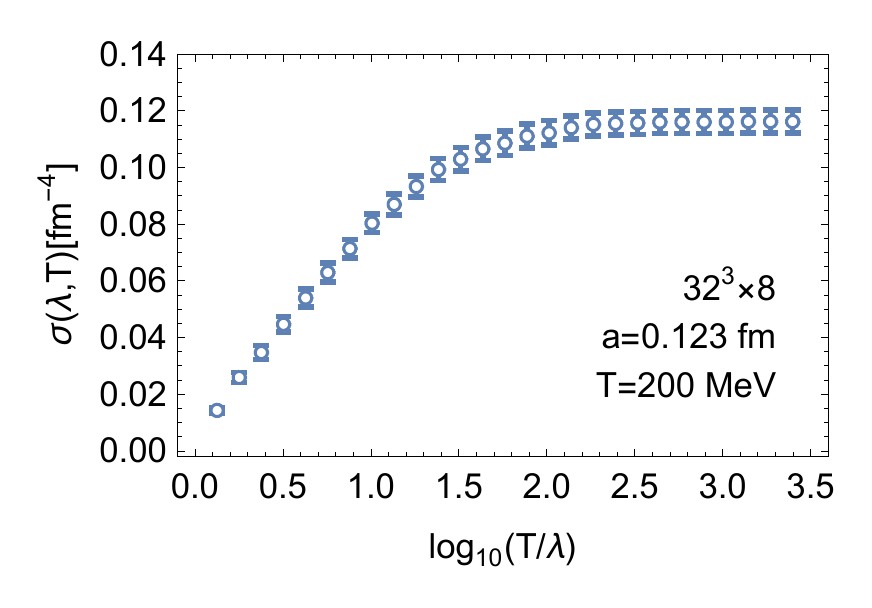}
    \hskip -0.08in
    \includegraphics[width=5.0truecm,angle=0]{sigmatoT_vs_lam_fodor_250MeV_nt8_n32.pdf}
    }
    \vskip -0.20in
    \caption{Thermal transition to the IR phase in QCD: $T_\fir$ is between $T\!=\!200\,$MeV and 
             $T\!=\!250\,$MeV.}
    \label{fig:IRcross2}
    \vskip -0.55in
\end{center}
\end{figure}

In addition to QCD data shown in Fig.~\ref{fig:IRcross} (bottom), we also 
computed the Dirac spectra at $T\!=\!200\,$MeV. The relevant comparison to 
spectral behavior in the IR phase (analog of Fig.~\ref{fig:IRcross}) is shown 
in Fig.~\ref{fig:IRcross2}. Note that $\sigma(x)$ is visibly approaching 
the linear regime that is characteristic of the IR phase. Together with 
the behavior of $\rho(\lambda)$ (left), this suggests that 
$T_\fir > 200\,$MeV, thus leading to the estimate \eqref{eq:030}.

\begin{figure}[t]
\begin{center}
    \centerline{
    \includegraphics[width=7.0truecm,angle=0]{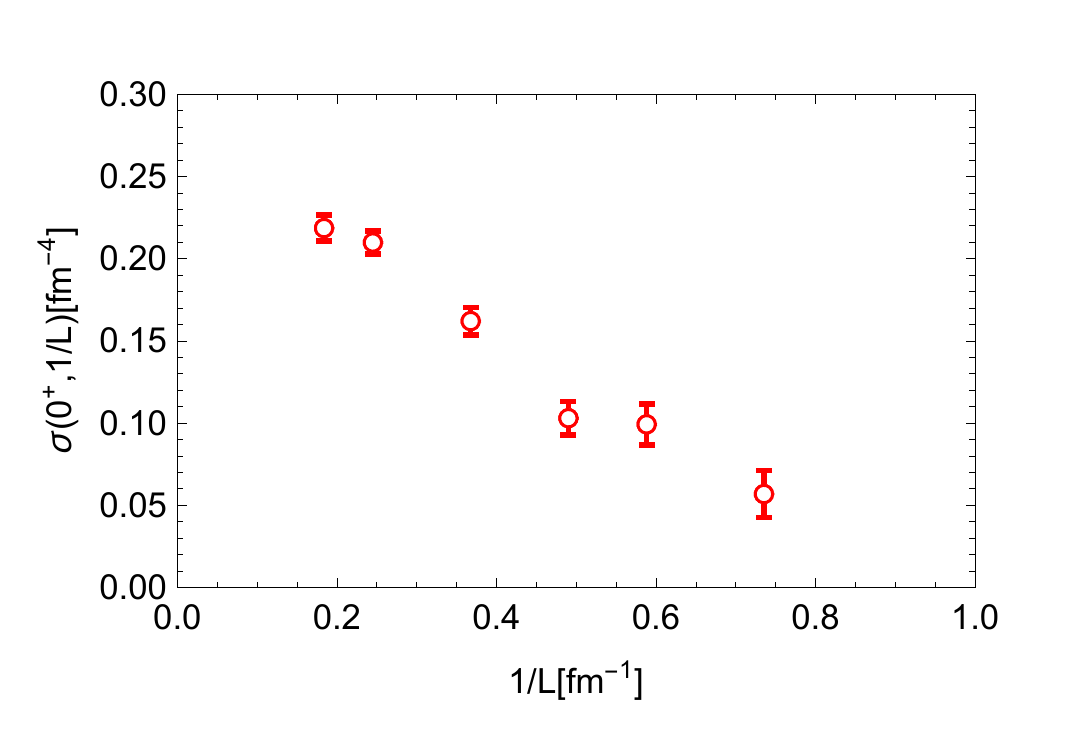}
    \hskip 0.10in
    \includegraphics[width=7.0truecm,angle=0]{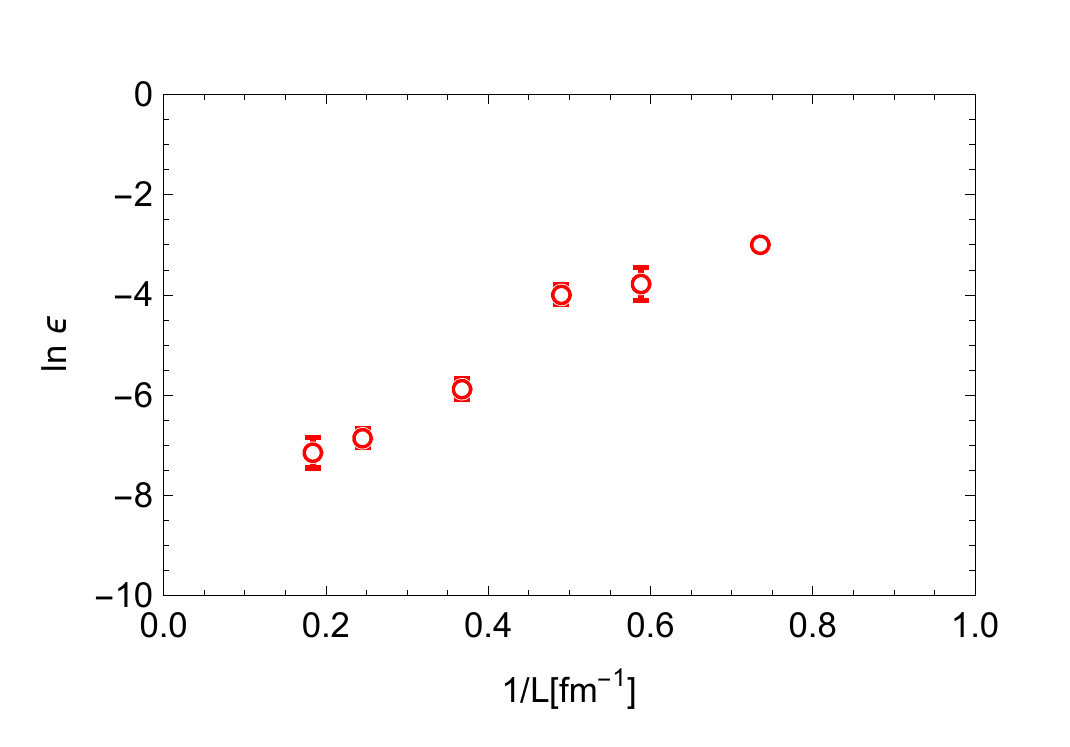}
    }
    \vskip -0.2in
    \caption{Left: the plot of $\sigma(0^+,1/L)$ for pgQCD ensembles. 
    Right: the plot of spectral edge parameter $\epsilon(L)$ (see text) for the same 
    ensembles.}
    \label{fig:supcond}
    \vskip -0.50in
\end{center}
\end{figure}

The scale $\Lambda_\fir \LessApprox T$ marks the upper edge of spectral region
where $\rho(\lambda) \propto 1/\lambda$ applies. For thermodynamic limit
considerations, it is desirable to ask whether our data also provides a hint of lower edge 
$\Lambda_\fir^{min}$ i.e. the point where the $1/\lambda$ either softens up, or 
the negative power behavior entirely disappears. A convenient indicator of this 
is $\sigma(0^+,1/L)$, namely the 4-volume density of non-zero Dirac modes smaller 
than IR cutoff $1/L$. Expressing the lower edge of Dirac spectrum at finite $L$ as 
$\epsilon(L)/L$, we have
\begin{equation}
    \sigma\left( 0^+,\frac{1}{L}\right) \,=\, 
    \sigma\left( \frac{\epsilon(L)}{L}, \frac{1}{L} \right) \,=\,
    \int_{\epsilon(L)/L}^{1/L} d \lambda \, \rho(\lambda)
    \label{eq:le010}
\end{equation}
which vanishes in $L \to \infty$ limit for all $p> -1$, where $\lambda^p$ is 
the leading IR behavior of $\rho(\lambda)$. Its $1/L$ behavior in our pgQCD 
ensembles is shown in Fig.~\ref{fig:supcond} (left). Since a turn toward zero for 
$1/L \to 0$ is not observed, the available data doesn't suggest the existence of 
$\Lambda_\fir^{min}$. 

Since the proportionality constant $c$ of $1/\lambda$ is stable with changing 
$L$ (see Fig.~\ref{fig:infrared}), and
\begin{equation}
    \lim_{L \to \infty}
    \sigma\left( 0^+,\frac{1}{L}\right) \quad \longrightarrow \quad
    -\lim_{L \to \infty} c(L) \ln \epsilon(L)
    \qquad \text{for} \qquad
    \rho(\lambda) = \frac{c}{\lambda} \quad , \quad 
    \lambda > \frac{\epsilon}{L} 
    \label{eq:le020}
\end{equation}
the finite $\lim_{L \to \infty} \sigma( 0^+,1/L) > 0$, conveyed by 
Fig.~\ref{fig:supcond} (left), implies finite  $\lim_{L \to \infty} \epsilon(L) > 0$. 
This is an important detail since $\epsilon(L)$ controls the IR edge of the spectrum. 
To check this explicitly, we estimate $\epsilon(L)$ directly using the relation
\begin{equation}
   \ln \epsilon(L) \,=\, 
   -k \, \frac{\sigma(0^+,1/L)}{\sigma(e^{-k}/L,1/L)} 
   \label{eq:le030}
\end{equation}
which holds for arbitrary $k \!>\! 0$ under the assumptions of~\eqref{eq:le020}. 
The result for $k\!=\!3$, suitable for our range of $1/L$ and the statistics, is 
shown in Fig.~\ref{fig:supcond} (right), confirming the trend toward small positive 
value of $\epsilon$ in $L \to \infty$ limit.

\end{appendix}

\bibliography{my-references}

\end{document}